\documentclass[prl,showpacs,twocolumn,floatfix]{revtex4}
\usepackage{amsmath}
\usepackage{graphicx}
\usepackage{placeins}

\begin{document}

\title{Dynamic Colloidal Stabilization by  Nanoparticle Halos}
\author{S. Karanikas}
\affiliation{Department of Chemistry, Lensfield Rd, Cambridge CB2 1EW, UK}
\author{A. A. Louis}\thanks{Author for correspondence: aal20@cam.ac.uk}
\affiliation{Department of Chemistry, Lensfield Rd, Cambridge CB2 1EW, UK}
\date{\today}
\begin{abstract}
We explore the conditions under which colloids can be stabilized by
the addition of smaller particles. The largest repulsive barriers
between colloids occur when the added particles repel each other with
soft interactions, leading to an accumulation near the colloid
surfaces. At lower densities these layers of mobile particles
(nanoparticle halos) result in stabilization, but when too many are
added, the interactions become attractive again.  We systematically
study these effects --accumulation repulsion, re-entrant attraction,
and bridging -- by accurate integral equation techniques.
\end{abstract}
\maketitle
\vspace*{-0.7cm}

Colloidal dispersions --~solid particles with radii ranging from a few
$nm$ to a few $\mu m$, suspended in a liquid solvent~-- are common in
nature and widely used in industry.  Blood, paint, ink and cement are
typical examples.  Because interatomic dispersion forces induce effective van
der Waals interactions with large attractive values at contact,
colloids will irreversibly aggregate, which is usually undesirable,
unless their surfaces are prevented from approaching too closely. The two
most common ways to achieve this are called {\em steric} and {\em
charge stabilization}\cite{Hunt01}.  Popular strategies for steric
stabilization usually involve grafting a layer of polymers onto the
colloid surface, resulting in dense repulsive brushes that prevent
close contact.
For charge stabilization, the route most common in
nature, the colloids have  surface charges of the same sign, leading to
a double layer of microscopic co- and counter-ions. Adding this 
effective repulsion to  the intrinsic van der Waals
attraction results in  the famous Derjaguin Landau Verwey Overbeek
(DLVO) potential\cite{DLVO}, with a metastable free-energy barrier
preventing aggregation.


In an important recent development, a third strategy for colloidal
stabilization, termed {\em nanoparticle haloing}, was introduced by
Lewis and co-workers\cite{Tohv01}.  By adding charged hydrous zirconia
nanoparticles of average radius $3 nm$ to a suspension of (marginally
charged) colloidal silica spheres of radius $285 nm$ in deionized
water, the following behavior was observed: For low nanoparticle
concentrations the silica spheres aggregate, driven by the generic van
der Waals attractions.  At intermediate nanoparticle concentrations,
the dispersion becomes stable, whereas at higher concentrations the
silica spheres aggregate again.  The authors\cite{Tohv01} attribute
the initial stabilization to layering of the small nanoparticles near
the colloidal surfaces.  These ``halos'' occur because it is
advantageous for the charged nanoparticles to be near the uncharged
colloid surfaces.  When two  colloids then approach each other,
their respective halos repel, preventing aggregation.  The
re-entrant aggregated phase, observed at higher nanoparticle
concentrations, was attributed to normal entropic depletion
attraction\cite{Asak58}.

Clearly, a new route to stabilize colloids would have many potential
applications. Indeed, nanoparticle haloing has already been used to
enhance the self-assembly of 3-D colloidal crystals on patterned
surfaces\cite{Lee04}.  Nevertheless, even though this novel
stabilization strategy has been demonstrated by experiment, many
questions remain about its generic applicability.  To address these
issues we carry out a systematic theoretical study of the effective
interaction $ \beta V_{bb}^{eff}(r)$\cite{Liko01} between colloids,
induced by (much) smaller particles ($\beta^{-1}$$=$$k_B T$ is the
reduced temperature.). We find  fairly large regime of parameter space where
$\beta V_{bb}^{eff}(r)$ is repulsive enough for stabilization, but this is
usually followed by re-entrant attraction at higher small particle
packing fraction.  The picture that emerges is considerably more
subtle than that of a simple static layer of adsorbed particles akin
to steric stabilization.  Instead, the nanoparticle halos are dilute,
and in dynamic equilibrium with the bulk solution.  Furthermore, we
observe no obvious change in their character when the re-entrant
attraction kicks in, implying that this phenomenon is more complex
than simple depletion attraction.

The key quantity we study is the effective interaction\cite{Liko01} $
\beta V_{bb}^{eff}(r)$ between two spheres of diameter $\sigma_{bb}$,
induced by smaller spheres.  Its properties are determined by the
number density $\rho_s$$=$$N_s/V$ of small particles and by the
interactions $\beta \Phi_{bs}(r)$ and $\beta \Phi_{ss}(r)$; 
it is independent of
the intrinsic interaction $\beta \Phi_{bb}(r)$\cite{Loui02a}.  If
$\Phi_{bb}(r)$ is attractive and leads to aggregation, then
introducing small particles that induce a $\beta V_{bb}^{eff}(r)$ repulsive
enough to counteract $\beta \Phi_{bb}(r)$ will stabilize the colloids.

The basic big-small and small-small interactions $\beta \Phi_{ij}(r)$ are
modeled by a hard-core Yukawa (HCY) form which is versatile without
having too many parameters to vary\cite{Loui02a}.
$\beta \Phi_{ij}(r) = \infty$ if $r$$<$$\sigma_{ij}$; 
$\beta\Phi_{ij}(r)=\phi_{ij}(r)$ for $r$$>$$\sigma_{ij}$, 
where in each case $r$ denotes the distance between the centers of the
particles, and the Yukawa tail is
 \begin{equation}\label{eq2.1}
\beta \phi_{ij}(r) = \frac{ \beta \epsilon_{ij} \sigma_{ij}}{r} \exp \left[ -
 \frac{(r - \sigma_{ij})}{\lambda_{ij}} \right],
\end{equation}
where $\sigma_{bs}$=$\frac12 (\sigma_{bb}$+$\sigma_{ss})$ with $\sigma_{ss}$ the small particle hard-core diameter.  By varying the size ratio
$q=\sigma_{ss}/\sigma_{bs}$, packing fraction $\eta_s=\frac16 \pi
\rho_s \sigma_{ss}^3$, and the 4 dimensionless potential parameters
$\beta \epsilon_{ss}$, $\beta \epsilon_{bs}$,
$\lambda_{ss}/\sigma_{ss}$ and $\lambda_{bs}/\sigma_{bs}$, a wide variety
of different physical situations can be studied\cite{Loui02a}. For
repulsive interactions, for example, Eq.~(\ref{eq2.1}) is a good model
for charged suspensions\cite{Hans00}.

\begin{figure}
\includegraphics*[width=8cm]{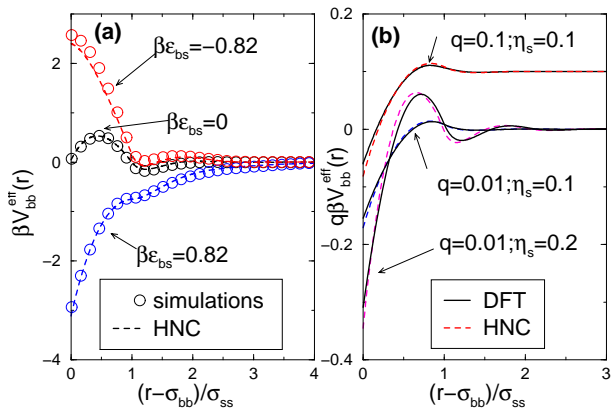}
\caption{\label{fig:hnc-simulations} (a): Comparison of
simulations (from \protect\cite{Loui02a})
and HNC calculations of the effective potential $\beta
V_{bb}^{eff}(r)$. The size-ratio $q=0.2$, and the potential parameters
$\lambda_{ss}/ \sigma_{ss} = 1/3$, $\lambda_{bs}/\sigma_{ss}=1/1.2$,
and $\beta \epsilon_{ss} = 2.99$ are kept constant, while $\beta
\epsilon_{bs}$ is varied.
 (b)~HNC and DFT\protect\cite{Roth00a} calculations for small
size-ratios $q$.  To facilitate comparisons, we plot $q\beta
V_{bs}^{eff}(r)$ and shift the curve for $q=0.1$ up by $+0.1$.
}\vspace*{-0.6cm}
\end{figure}

To restrict this vast parameter space somewhat, and inspired by the
 successful experiments\cite{Tohv01}, we choose $\epsilon_{ss} >0$
and, initially, $\epsilon_{bs} =0$. The effective
potentials $\beta V_{bb}^{eff}(r)$ are calculated by using the two-component Ornstein
Zernike (OZ) equations in the $\rho_b\rightarrow 0$ limit where they
decouple, together with the hypernetted chain (HNC) integral equation
closure\cite{Hans86}, leading to: $\beta V_{bb}^{eff}(r) = -\rho_s
\int d{\bf r'} h_{bs}(r) c_{bs}(|{\bf r'} - {\bf r}|)$, where
$h_{bs}(r) = g_{bs}(r)-1$ with $g_{bs}(r)$ the pair correlation
function between big and small particles, and $c_{bs}(r)$ is the
direct correlation function, related to $h_{bs}(r)$ by the OZ~equation\cite{Hans86}.  In this limit, HNC has some important
advantages\cite{Kara04} over other popular integral equations such as Percus Yevick
(PY) or Rogers Young (RY)\cite{Hans86}. For example, it is~exact for
the AO model\cite{Asak58} at all densities $\rho_s$ (PY is not\cite{Loui01a,Kara04}).  Moreover, HNC is known to be particularly accurate
for soft repulsive potentials of the type we are
investigating\cite{Hans86}.  To validate our method, we compare, in
Fig.~\ref{fig:hnc-simulations}, the performance of HNC with
several simulations\cite{Loui02a} for $\epsilon_{ss} >0$, and find
excellent agreement.  Since we also want to study rather extreme size
ratios, we compare, in Fig.~\ref{fig:hnc-simulations}b, to depletion
potentials for hard spheres (HS) calculated with an accurate Density
Functional Theory (DFT) approach\cite{Roth00a}. Again HNC performs
remarkably well.  These results provide the confidence that, even if
HNC is not perfectly quantitative, the trends we uncover will be
robust, provided  we limit ourselves to soft repulsions and
 low packing fractions\cite{HNC}. Fortuitously, this appears
to be the regime where the nanoparticle haloing mechanism operates
most effectively.

\begin{figure}
\includegraphics*[width=8cm]{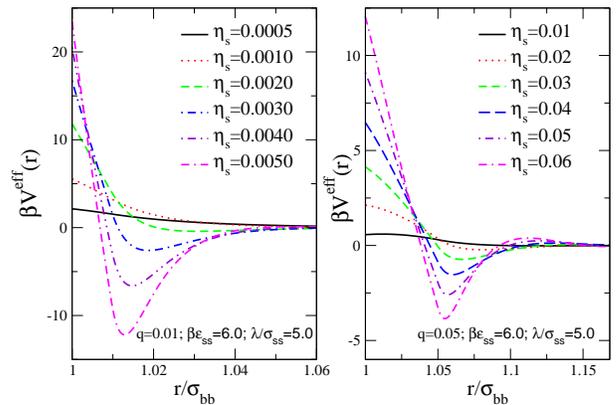}
\caption{\label{fig:effective-potential} Two typical examples
of the  effective potentials. First there is initial stabilization, and then a
 re-entrant attraction.  }\vspace*{-0.6cm}
\end{figure}

In order to systematically investigate the conditions for which
repulsive stabilization occurs, we calculated, with HNC, $\beta
V^{eff}_{bb}(r)$ for a large number of parameter combinations.  Two
typical examples are shown in Fig.~\ref{fig:effective-potential},
demonstrating the common pattern we find: for increasing packing  fractions a
maximum first appears close to contact, and  continues to increase until
at
higher $\eta_s$ a  secondary minimum appears that grows
with $\eta_s$ and rapidly moves to a separation of about one
$\sigma_{ss}$.  This sequence of initial stabilization followed by
re-entrant attraction  is similar to that seen in the
experiments\cite{Tohv01}, and is found throughout the parameter regime we
investigated.

\begin{figure}
\includegraphics*[width=8cm]{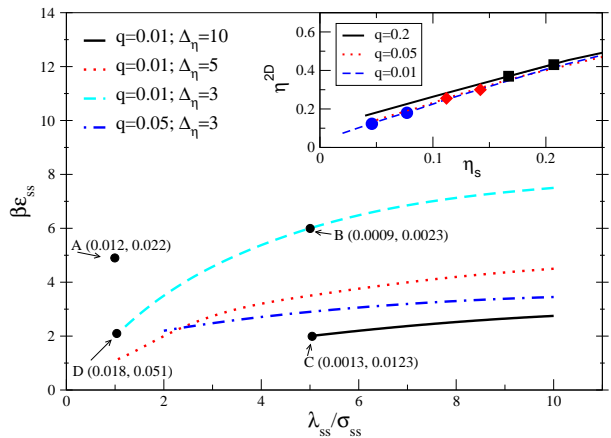}
\caption{\label{fig:epsvslambda}
 The ``equi-$\Delta_\eta$'' lines denote different values of
the stability window $\Delta_\eta = \eta_s^u/\eta_s^s$ for size-ratios
$q=0.01$ and $q=0.05$.
 For points A-D the stability window packing
fractions, for $q=0.01$, are listed in the format ($\eta_s^s,\eta_s^u$).
Inset: The 2D packing fraction $\eta^{2D}$ of a ``halo''  for different
size-ratios $q$. The  potential parameters are $\beta \epsilon_{ss}=2$, and
$\lambda_{ss}/\sigma_{ss}=0.5$. The symbols denote bulk packing fractions $\eta_s^s$ and $\eta_s^u$ for $q=0.01$
(circles), $q=0.05$ (diamonds) and $q=0.1$ (squares). }
\vspace*{-0.6cm}
\end{figure}

 To further quantify the region of stability, we choose the following
 measure: For a given set $\beta \epsilon_{ss}$,
 $\lambda_{ss}/\sigma_{ss}$, and $q$, we calculate the effective
 potentials for different $\eta_s$, as done for
 Fig.~\ref{fig:effective-potential}. The ``stability window'' is
 defined as $\Delta_\eta = \eta_s^u/\eta_s^s$, where $\eta_s^s$ is the
 packing fraction above which the maximum of $\beta V_{bb}^{eff}(r)$
 is $>5$ (leading to kinetic stabilization), and $\eta_s^u$ is the
 packing fraction below which the minimum of $\beta V_{bb}^{eff}(r)$
 is $< -2$ (a conservative estimate of where short-range attractions
 induce aggregation\cite{Loui01a}).  The way our stability measure
 $\Delta_\eta$ varies with potential parameters is depicted in
 Fig.~\ref{fig:epsvslambda}, from which some general trends can be
 extracted: The size of the window increases with increasing
 $\lambda_{ss}/\sigma_{ss}$ and decreasing $\epsilon_{ss}$ and $q$.
 (In the HS limit ($\beta \epsilon_{ss}=0$) we find no window of
 stability).  For a number of points (A-D) in
 Fig.~\ref{fig:epsvslambda}, we show the values of $\eta_s^u$ and
 $\eta_s^u$.  As expected, these packing fractions decrease with
 increasing $\beta \epsilon_{ss}$ and $\lambda_{ss}/\sigma_{ss}$ since
 the small particles repel each other more and have a larger effective
 ``size''. 

The effective repulsion is clearly related to the accumulation of
particles near the colloids (nano-particle halos\cite{Tohv01}).  We
define the ``halo'', as those particles between $r=\sigma_{bs}$ and
$r=r_{min}$, the distance at which the pair correlation function
$g_{bs}(r)$ has its first minimum. The number of particles $N_{halo}$
follows from integrating $g_{bs}(r)$ up to $r_{min}$.  The 2-D packing
fraction is given by $\eta^{2D}=\frac14 \pi \rho^{2D} \sigma_{ss}^2$
where $\rho^{2D} = N_{halo}/( 4 \pi \sigma_{bs}^2)$.  For all the
parameters studied we find the same behavior depicted in the inset of
Fig.~\ref{fig:epsvslambda}: The halo packing fraction $\eta^{2D}$ is
linear with $\eta_s$ and there is no change of slope or other obvious
property marking either the beginning of stabilization at $\eta_s^s$
or re-entrant attraction at $\eta_s^u$.  We have also investigated
other surface properties such as the adsorption $\Gamma_s = -\int
h_{bs}(r) d{\bf r}$ and the related surface tension
$\gamma_s$\cite{Loui02b}. In contrast to polymeric depletants, where
$\gamma_s$ helps determine $\beta V_{bb}^{eff}(r)$\cite{Loui02b}, we
observe no clear signatures of $\eta_s^s$ or $\eta_s^u$ in the surface
tension or the adsorption..

 For small $q$ we expect the
$g_{bs}(r)$ to be similar within corrections ${\cal O}(q^{-1})$,
which explains why the $\eta^{2D}$ v.s.\ $\eta_s$ curves are so close
for different $q$.  However, the stability windows, shown by the
symbols in the inset of Fig.~\ref{fig:epsvslambda}, differ
significantly: They are at lower $\eta_s$ for smaller $q$, something we
observe more generally.  This can be understood from an approximate
Derjaguin\cite{Hunt01} argument valid in the small $q$ limit.  The
potentials scale as $1/q$ times the force between two plates, and so
stabilization (and re-entrant attraction) are achieved at lower
packing fractions.  However, this doesn't easily explain why the
window size also grows with decreasing $q$.

The halos are very dilute at stabilization, and we have checked that
all layer densities studied are well below that of any
two-dimensional freezing transition.  In fact, the layers shown in the
inset of Fig.~\ref{fig:epsvslambda} are among the densest we
investigated; for some $\Phi_{ss}$, $\eta_s^{2D}$ can easily be an
order of magnitude lower at $\eta_s^s$.  At these low packing
fractions, the particles rapidly diffuse between halos and the bulk.
In contrast to a steric stabilization mechanism, where the layers are
static, we emphasize that this nanoparticle halo stabilization
mechanism is {\em dynamic}.

Further evidence against a naive picture of static layers comes from
the re-entrant attraction.  If the halos would become saturated, so
that additional small particles can no longer segregate to the
colloidal surface and instead act as depletants, then one might expect
a linear dependence of the minimum of $\beta V_{bb}^{eff}(r)$ on
$\eta_s$ as in AO\cite{Asak58} or HS\cite{Roth00a} depletion.  Instead,
the minimum in $\beta V_{bb}(r)$ grows initially as $\eta_s^2$, 
closely resembling the behavior of the second minimum of HS
systems\cite{Roth00a}, which  suggests that both minima have a similar more
complex origin in correlation effects.  In fact, they would be
directly related if the potentials were interpreted in terms of a
non-additive HS reference system with $\sigma_{bs} <
\frac12(\sigma_{bb}+\sigma_{ss})$, as explained in \cite{Loui01b}.
  The repulsive
effective interactions found in many other theoretical studies of
$\beta V_{bb}^{eff}(r)$ can also be qualitatively interpreted in this
way (see e.g.\ \cite{Loui02a} and references therein for a
discussion), suggesting that non-additivity may  be fruitfully used to
interpret the re-entrant attraction\cite{Kara04}.

\begin{figure}
\includegraphics*[width=8cm]{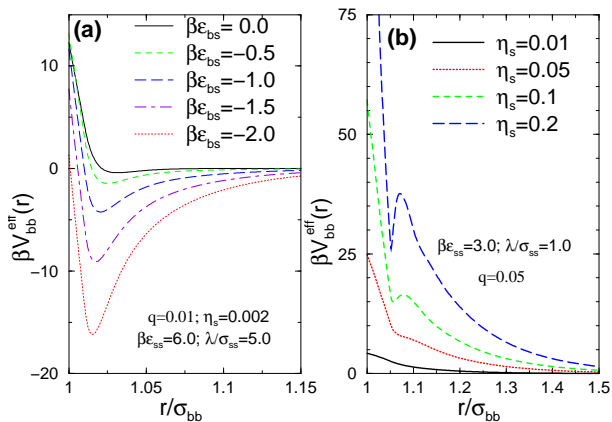}
\caption{\label{fig:bridging} (a) Adding an attractive $\beta \phi_{bs}(r)$ generates a deep minimum due to bridging effects
when $\lambda_{bs}=\lambda_{ss}$.
 (b) A dramatic stabilization effect occurs for weak longer ranged
attractions: $\lambda_{bs} = 3 \lambda_{ss} = 3 \sigma_{ss}$, and
$\beta \epsilon_{bs}=-0.5$.  }\vspace*{-0.6cm}
\end{figure}

One might argue that since adding an attractive $\phi_{bs}(r)$ should
increase the number of particles in a layer, this should enhance the
stabilization effect.  However, we find  more subtle scenarios.  If
we choose $\lambda_{bs}=\lambda_{ss}$, to model residual charge on the
large colloids, then for weak attractions the window indeed grows
slightly. But, as $\beta\epsilon_{bs}$ becomes more negative, the
potentials rapidly develop a large attractive component.  This
phenomenon, sometimes called {\em bridging} for polymeric
additives\cite{Hunt01}, results from configurations where the two
bigger colloids are both attracted to the same set of smaller
particles\cite{Kara04}.  An example of bridging is demonstrated in
Fig.~\ref{fig:bridging}(a), and is representative of what we find more
generally:  the stability
window $\Delta_\eta$ initially grows slightly, but then rapidly
disappears, typically around $\beta \epsilon_{bs} \lesssim -1.5$.

On the other hand, a dramatic enhancement of the stabilization occurs
for longer ranged colloid-nanoparticle attractions, as demonstrated in
Fig.~\ref{fig:bridging} for $\lambda_{bs} = 3 \lambda_{ss}$. The
bridging effect is bypassed and the first minimum shifts up to
positive absolute values, in fact for these parameters we find no
re-entrant attraction within the range where we trust HNC\cite{HNC}.
In general we find this effect for $\lambda_{bs} > \lambda_{ss}$, but
exactly where it kicks in depends $q$ and the other potential
parameters. In all cases studied, the 2D layer densities are still
very low so that the ``halos'' are dilute; typically for more negative $\beta
\epsilon_{bs}$ bridging sets in again\cite{Kara04}.  Of course when
$\beta \phi_{bs}(r)$ is attractive  enough to induce static saturated layers,
then the particles would be sterically stabilized.  But technically
this is a non-equilibrium effect: it only works if the colloidal
particles are first isolated from each other on the timescale that the
(saturated) layer forms.

Finally, having systematically explored the effect of different
parameters on $\beta V_{bb}^{eff}(r)$, we make some recommendations
for experiments.  For $\epsilon_{bs}=0$ the best stabilization should
occur for modestly charged small additives (nanoparticles) since the
stability windows are largest for small $q$, modest $\beta
\epsilon_{ss}$ and large $\lambda_{ss}/\sigma_{ss}$. Small
$\sigma_{ss}$ are needed to enhance the latter parameter, since the
(Debye) screening length is typically fixed by solution conditions,
e.g.\ $\lambda_{ss} \approx 30 nm$ for $0.1mM$ monovalent salt
concentration.  Another argument in favor of small particles concerns
the {\em dynamic} nature of the layers.  If the big particles are
driven at velocities such that the smaller particles can no longer
adiabatically follow, then the stabilization effect may
disappear. Since the self-diffusion coefficients of the nanoparticles
scale as $1/\sigma_{ss}$\cite{Hunt01}, this again favors small
particles.  Moreover, larger windows $\Delta_\eta$ also enhance
stability under halo fluctuations.

 Whereas adding charge to the colloids can destroy the stabilization
effect,  a modest but longer ranged attraction $\beta
\phi_{bs}(r)$ can significantly enhance it.  The latter effect could
be induced by residual van der Waals attractions, although this
recommendation must be tempered by the difficulty of adding van der
Waals attractions between different species without simultaneously
significantly increasing them between similar species.  On the other hand, the
advantage of small van der Waals attractions (which are independent of
Debye screening length), is that the ratio $\lambda_{bs}/\lambda_{ss}$
can be tuned by changing salt concentration.  This adds another handle
for engineering  effective potentials and concomitant phase
behavior\cite{Kara04}.

We observe the same general trends seen in  the experiments of
ref.~\cite{Tohv01}, such as lower
$\eta_s^s$ with smaller $q$ and values of $\eta^{2D}$ well below
saturation. A direct quantitative comparison, however,  
 is  hampered by their use of polydisperse
small particles, and the difficulty in deriving accurate potentials
$\beta\phi_{ij}(r)$.  Taking their estimates($\beta \epsilon_{ss}
\approx 6$, $\lambda_{ss}/\sigma_{ss} \approx 0.6$) we find a smaller
window $\Delta_\eta \approx 2$, at higher $\eta_s$ than what they
observed. The difference could stem  from a small
attractive $\beta \phi_{bs}(r)$ or from  polydispersity (preliminary
calculations suggest that this lowers the effective $\eta_s^s$ and
$\eta_s^u$ (P. Bryk, {\em private communication})).



In conclusion, we discovered a substantial parameter regime where the
addition of small (nano) particles can stabilize bigger
colloids. Fortuitously, this occurs where the flexible HNC integral
equation is most reliable. We usually find a stability window of
packing fractions, above or below which the colloids aggregate
again. The effects are significantly enhanced for weak longer ranged
attractive $\beta \phi_{bs}(r)$.  Although the stabilization is
clearly related to the formation of diffusive accumulation layers
around the bigger particles, we find no simple relationship to layer
properties. This suggests these effects are related to more complex
correlations. Colloidal stabilization by dynamic nanoparticle halos
should be widely applicable and complimentary to existing steric and
charge stabilization techniques\cite{Hunt01,DLVO}.  This new mechanism
may also have relevance to smaller scale biological
interactions\cite{Leck01}.  [Note: upon completion of this work we
became aware of a study by J. Liu and E. Luijten, cond-mat/0411278,
which uses different techniques, but arrives at similar conclusions]


We thank H L\"{o}wen for early discussions, and J. Dzubiella, R. Roth
and P. Bryk for
invaluable help with the calculations.  SK thanks Schlumberger Cambridge
Research and the EPSRC for a studentship, and AAL thanks the Royal Society (London) for financial support.

\FloatBarrier


\begin{thebibliography}{99}
\bibitem{HNC} We stress that no claim is being made for the
general accuracy of HNC for all parameter values.  For example 
the predictions for HS begin to show important quantitative
deviations for $\eta_s > 0.3$ and  the comparison with simulations
of ref~\protect\cite{Loui02a} is less accurate for
attractions $(\epsilon_{ss} < 0)$.
\end{thebibliography}
\end{document}